\documentclass[fleqn,twoside]{article}
\usepackage[headings]{espcrc2}

\readRCS
$Id: espcrc2.tex,v 1.2 2004/02/24 11:22:11 spepping Exp $
\usepackage{epsfig}
\usepackage[figuresright]{rotating}

\title{
\vspace{-2cm}
\hfill \parbox{2cm}{\normalsize \tt IFIC/07-02} \vspace{1.5cm} \\
PHOKHARA, the radiative return and the $(g-2)_{\mu}$ puzzle}

\author{Germ\'an Rodrigo~\address[IFIC]
       {Instituto de F\'{i}sica Corpuscular, 
        CSIC-Universitat de Val\`encia, \\
        Apartado de Correos 22085, E-46071 Valencia, Spain \\
        E-mail: german.rodrigo@ific.uv.es}
        \thanks{Work supported by Ministerio de
Educaci\'on y Ciencia (MEC) under grant FPA2004-00996, 
Generalitat Valenciana under grant GV05-015,
Consejo Superior de Investigaciones Cient\'{\i}ficas (CSIC) 
under grant PIE 200650I247, and European Commission FLAVIAnet 
MRTN-CT-2006-035482.}}
       
\runtitle{}
\runauthor{}

\begin{document}

\begin{abstract}
The radiative return has proven to be a competitive method for 
the precise measurement of the hadronic cross section, detailed 
studies of hadronic interactions, and even discoveries of new 
resonances. The most recent and future devolopments of the 
Monte Carlo event generator PHOKHARA are highlighted, 
and the impact of the radiative return measurements on the
$(g-2)_\mu$ puzzle is discussed. 
\end{abstract}

\maketitle

\section{Introduction}

Electron--positron annihilation into hadrons is one of the basic
reactions of particle physics, relevant for the understanding
of hadronic interactions. The low energy region is crucial
for predictions of the hadronic contributions to $a_{\mu}=(g-2)_{\mu}/2$, 
the anomalous magnetic moment of the muon, and to the running of the 
electromagnetic coupling from its value at low energy up to $M_Z$.

The traditional way of measuring the hadronic cross section, the 
energy scan, needs dedicated experiments. 
An alternative and advantageous way is the radiative return method. 
This method allows for a simultaneous extraction of the hadronic cross 
section over a wide energy range in an homogeneous data set, 
profiting from the data of all high luminosity meson factories.

\begin{figure}[htb]
\begin{center}
\epsfig{file=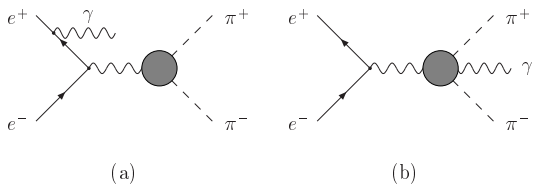,width=7.5cm} \vspace{-.7cm}
\caption{Leading order contributions to the reaction 
$e^+e^-\to\pi^+\pi^-\gamma$ from ISR~(a) and FSR~(b).}
\label{fig1}
\end{center}
\end{figure}

The radiative return method relies on the observation that the cross section 
of the reaction $e^+e^- \to$ hadrons+photons, with photons emitted from the
initial leptons (ISR), see Fig.~\ref{fig1}(a), 
factorizes into a function $H$, fully calculable within QED,
and the cross section of the reaction $e^+e^- \to$ hadrons.
Thus from the measured differential cross section 
of the reaction $e^+e^- \to$ hadrons + photons 
as a function of the hadronic invariant mass one can evaluate 
$\sigma(e^+e^- \to$ hadrons$)$ once the radiator function $H$ is known. 
The radiative return method allows for 
the extraction of the hadronic cross section from the production energy 
threshold of a given hadronic channel almost to the nominal energy of 
the experiment. The smaller cross section of the radiative process 
as compared to the process without photon emission has to be 
compensated by higher luminosities. 
That requirement is met by meson factories (DAPHNE, CLEO, BABAR, BELLE). 
All of them were built for other purposes than the measurement of the 
hadronic cross section, but their huge luminosities provide with data 
samples large enough for very accurate measurements of interesting hadronic
channels and/or give information on rare channels, which were not 
accessible in scan experiments.

Two representative examples of such measurements are the very accurate 
pion form factor extraction by the KLOE Collaboration~\cite{KLOElarge}, 
and the measurement of $\sigma(e^+e^- \to$ 3 pions$)$ by the BABAR 
Collaboration~\cite{Aubert:2004kj}, where it was shown that the old DM2 
scan data were too low at high values of the invariant mass of the 
hadronic system. For a review of BABAR results see Ref.~\cite{wang}.

\section{The PHOKHARA event generator}

In realistic experimental situations, where sophisticated event selections 
are used, one needs a Monte Carlo event generator of the measured process. 
To meet that requirement the PHOKHARA event generator 
(http://cern.ch/german.rodrigo/phokhara/\footnote{Mirror link 
at http://ific.uv.es/$\sim$rodrigo/phokhara/}) was constructed. PHOKHARA 
started from the EVA generator~\cite{Binner:1999bt}, 
where the structure function method was used to model multi-photon 
emission. The physical accuracy of EVA 
was however far from the demanding experimental accuracy for the 
measurement of the pion form factor. PHOKHARA relies instead on exact 
matrix elements at next-to-leading order (NLO), namely it includes 
one loop radiative corrections to one-photon radiation and emission 
of two real hard photons. The accuracy of the simulation 
has been estimated to be of the order of 5 per mil 
from ISR~\cite{Rodrigo:2001kf}. 

The first version of PHOKHARA~\cite{Rodrigo:2001kf}
was designed to run with tagged photon configurations. 
Radiative corrections necessary for photon emission 
at small angles were calculated afterwords~\cite{Kuhn:2002xg}
and implemented into the event generator~\cite{Czyz:2002np}. 
The important issue of final state emission was addressed 
subsequently~\cite{Czyz:2003ue}. 
In parallel the generator was being extended to allow for 
the generation of more hadronic channels. The present version of the 
program simulates the production of not only a pair of pions or muons, 
but includes also the simulation of $K^+K^-$, $\bar{K}^0K^0$, events with 
three~\cite{Czyz:2005as} and four pions~\cite{Czyz:2002np}, 
and nucleon pairs $p\bar{p}$ and $n\bar{n}$ ~\cite{Czyz:2004ua} . 

All that allowed for building the state-of-the art event generator. 
The proper implementation of the radiative corrections as well as the 
hadronic currents is guarantied by extensive tests. 
Comparisons with the KKMC~\cite{Jadach:1999vf} Monte Carlo 
event generator have been performed~\cite{Jadach:KKMC} 
leading to an excellent agreement. The comparison is however limited 
to muons in the final state. Higher order effects, that can be seen 
as a difference between exponentiated and non-exponentiated 
matrix elements reach at most 2 per mile with the exception of the 
region where the hadronic system has an invariant mass very close to 
the nominal energy of the experiment. There, soft multi-photon 
emission plays an important role and thus exponentiation is necessary.
This region is however out of the region of interest for  
radiative return measurements~\cite{Rodrigo:2001kf}.

The impresive amount of new data provided in particular by $B$-factories 
requires further efforts to 
improuve the accuracy of the event generator, as well as to implement new 
hadronics channels. The latter requires a fairly good parametrization of 
various form factors.

\section{Final state radiation and radiative $\phi$-decays} 

Final state radiation (FSR), see Fig.~\ref{fig1}(b), is the 
main background for radiative return measurements. 
The situation at $B$-factories is however completely different 
from the one at the $\phi$-factory DAPHNE. 
In the former case the region of hadronic masses below 4~GeV, 
which is of the utmost physical interests, lays far from the nominal 
energy of the experiments. Thus an emission of a very hard photon 
is required to reach it. As a result the typical kinematic configuration 
of an event consists of a photon emitted back-to-back to the 
hadronic system. That provides a natural suppression of
FSR contributions, which are large for photons emitted parallel
to the direction of the charged hadrons in the final state. 

At the $\phi$-factory the physically interesting region is not so far 
from the nominal energy of the experiment, and that natural suppression  
of FSR do not hold. Strategies should be stablished to either suppress 
FSR through kinematical cuts, or to control the uncertainty due to 
the model dependence of the simulation. 
The KLOE analysis~\cite{KLOElarge,venanzoni} of the pion form factor from 
events with untagged photons emitted at small polar angles follows the 
first strategy. When photons in the forward-backward 
directions and pions in the central region are selected 
FSR is easily reduced to less than 1\%. The price to pay however 
is that the region close to threshold, $M_{\pi\pi}<590$~MeV, 
is also suppressed as pions are produced in this kinematical region 
essentially back-to-back to the ISR photon and therefore at very 
small polar angles that scape from detection. 
For the case of untagged photons a specific background,
$e^+e^-\to\pi^+\pi^-e^+e^-$, has to be also taken into 
account~\cite{Hoefer:2001mx,Czyz:2006dm} as the final leptons are not vetoed.

KLOE is now perfoming a complementary analysis with photons tagged 
at large polar angles~\cite{venanzoni} that will provide accurate data in 
the threshold region. Due to the factor $1/s^2$ of the dispersion 
integral for $a_\mu$, that low energy region is highly enhanced; 
contributing to around $20\%$ to the total integral. Therefore its
relevance. 

At large photon polar angles FSR and $\phi$ decays: 
$\phi \to \pi^+\pi^-\pi^0$ and $\phi \to f_0 \gamma
\to \pi^+\pi^-\gamma$~\cite{Melnikov:2000gs}; become important.
The background channel $\phi \to \pi^+\pi^-\pi^0$
can be eliminated through dedicated selection cuts,
but FSR and $\phi \to \pi^+\pi^-\gamma$ have to be
subtractred relying on Monte Carlo. Another possibility 
to eliminate $\phi$ decays is to run off-peak.
DAPHNE has taken data off-resonance at a center-of-mass
energy of $\sqrt{s}=1$~GeV that will allow to reduce the 
systematic errors in the threshold region~\cite{venanzoni}.

The main tool to test the model dependence of photon
emission from the final state pions and radiative 
$\phi$-decays is the charge asymmetry.
For events with emission of one real photon the two-pion state is produced 
with charge conjugation $C=-1$ and odd orbital angular momentum
when the real photon is emitted from the initial state, 
and with $C=+1$ and even orbital angular momentum when the real 
photon is emitted from the final state. 
As a result, the ISR-FSR interference is odd under 
the exchange $\pi^+\leftrightarrow \pi^-$ 
and integrates to zero for charge blind event selections. 
At the same time it is the only source of the charge asymmetry 
and as such allow to test the FSR model. 

As shown in~\cite{Czyz:2004nq}, the charge asymmetry has large analyzing power 
and can provide information allowing for distinguishing between different models
of the radiative $\phi\to\pi\pi\gamma$ decay~\cite{KLOEphi}.
PHOKHARA have adopted two models describing the decays $\phi \to \pi^+ \pi^-
\gamma$ and $\phi \to \pi^0 \pi^0 \gamma$: 
the ``no structure'' model~\cite{Bramon:1992ew}
and the $K^+K^-$ model~\cite{LucioMartinez:1994yu}.
Again by appropriate event selections one can suppress those contributions 
or enhance them as for other sources of FSR emission.
Other contributions that might be important in the threshold region 
beyond the model currently used in PHOKHARA to describe FSR 
(sQED + vector dominance + radiative $\phi$ decays) have been advocated 
in Ref.~\cite{Pancheri:2006cp}.

The reaction $e^+e^-\to \pi^+\pi^-\gamma$,
with the photon emitted from the pions, does contribute also to 
dispersion integrals for the evaluation of $a_\mu$ and $\alpha_{QED}$.
In the former case its theoretically estimated value~\cite{Czyz:2003ue}
is of the size of the theoretical uncertainty and thus numerically
important. As its theoretical estimations are not reliable it has 
to be measured. PHOKHARA also includes the simulation of events
with simultaneous emission of one photon from the initial state 
and another from the final state.

\section{Three-pion channel}

The channel $e^+e^- \to \pi^+ \pi^- \pi^0$ has been recently 
implemented in PHOKHARA~\cite{Czyz:2005as}. The model for the form
factor is based on generalized vector dominance with isospin $I=0,1$ 
components:
\begin{eqnarray}
&& \!\!\!\!\!\!\!
J_\mu^{em,3\pi} = \langle \pi^+(q_+)\pi^-(q_+)\pi^0(q_0) \mid
J_\mu^{em} \mid 0 \rangle \nonumber \\ 
&& = \epsilon_{\mu\alpha\beta\gamma} q_+^\alpha q_-^\beta q_0^\gamma
\sum F_{3\pi}^{I=0,1}
(q_+,q_-,q_0)~. 
\end{eqnarray}
A global fit has been performed with contributions from
$\omega(782)$, $\omega'=\omega(1420)$, $\omega''=\omega(1650)$,
$\phi(1020)$, $\rho(770)$, $\rho'=\rho(1450)$
and $\rho''=\rho(1700)$ resonances. The model provides a very 
good description of the total cross section and also predicts 
in good agreement with experiment the decay width 
$\Gamma(\pi^0 \to \gamma \gamma)$, 
the slope parameter of $\pi^0 \to \gamma \gamma^*$, and the 
radiative vector meson decays $\rho \to \pi^0 \gamma$ and  
$\phi \to \pi^0 \gamma$, but is however in conflict with  
$\omega \to \pi^0 \gamma$.  There is still room from improvements 
once information on subdistributions is included in the fit. 

\section{Nucleon form factors}
  
Another example of the power of the radiative return method
is the measurement of the magnetic and the electric nucleon 
form factors in the time-like region. 
This measurement was first proposed in Ref.~\cite{Czyz:2004ua}. 
The recent measurement by the BABAR Collaboration~\cite{Aubert:2005cb} 
of the magnetic and electric proton form factor shows a clear 
evidence for a ratio $|G_E/G_M|>1$ just above threshold
with respect to to previous analysis.
It is also interesting to note
that the measurement of the relative fase between $G_E$ and $G_M$ 
requires access to the nucleon spin~\cite{Czyz:2004ua} .

\section{Production and decay of $J/\psi$ and other narrow resonances}

$J/\psi$ resonances are copiously produced at B-factories. 
There is a strong demand from experimental groups to include 
this kind a new channels in our program, to analyse its reach 
phenomenology with high precision. 
Other narrow resonances are of interest as well. 

The BABAR Collaboration~\cite{Aubert:2005rm} has reported the 
discovery of a new state, the Y(4260) resonance, by using radiative 
return events in the $\pi^+ \pi^- J/\psi$ channel. These results have 
been confirmed by CLEO using energy-scan~\cite{Coan:2006rv},
and ISR~\cite{He:2006kg} data, as well as by the BELLE 
Collaboration~\cite{Abe:2006hf}.

\section{One-loop corrections to muon production and higher order 
radiative corrections}

When the first version of PHOKHARA was constructed the two-pion channel 
was the most interesting as it enters the prediction for the anomalous 
magnetic moment of the muon. By that time ISR radiative 
corrections where introduced in PHOKHARA through a leptonic tensor:
\begin{eqnarray}
&& \!\!\!\!\!\!\!\!
L_{\rm{ISR}}^{\mu\nu} = \alpha^2 \left[ a_{00} \, g^{\mu\nu}
+ a_{11} \, p_1^\mu p_1^\nu + a_{22} \, p_2^\mu p_2^\nu \right. \nonumber \\
&& \left. + a_{12} (p_1^\mu p_2^\nu+p_2^\mu p_1^\nu)
+ i \pi a_{-1} (p_1^\mu p_2^\nu-p_2^\mu p_1^\nu) \right]~, \nonumber \\
\end{eqnarray}
where $p_1$ ($p_2$) are the four-momentum of the incoming
positron (electron), and $a_{ij}$ are scalar coefficients 
where the antisymmetric imaginary contribution proportional 
to $a_{-1}$ appears first at NLO. 
While ISR corrections are independent of the hadronic channel, FSR
and the ISR-FSR interference do depend and have 
to be calculated for each channel independently. 

When more and more hadronic channels are of interest it seems more 
convenient to introduce radiative corrections at the amplitude level by 
using the helicity amplitude formalism. Interferences between different
amplitudes are then obtained automatically without further 
analytical computations. The radiative one-loop amplitude can be 
factorized into three components:
\begin{equation}
| A \rangle = | A \rangle_{\rm{ISR}} + | A \rangle_{\rm{FSR}}
+ | A \rangle_{2\gamma^*}~,
\end{equation}
where the last one steams from the exchange of two virtual photons 
between the initial and the final state. 

Muon pair production is not only important for the normalization of 
the $R$-ratio, but being a very clean process that can be calculated 
in QED can be used for luminosity monitoring at $e^+e^-$ machines. 
Therefore the importance of having very accurate predictions. 
We are now completing the calculation of ISR radiative corrections 
at the amplitude level and FSR for the muon channel. The 
two-photon exchange amplitude will be calculated subsequently.
Note that the latter is not even known for Bhabha scattering. 

The estimated accuracy of PHOKHARA from ISR is of the order of 5 per mil. 
Although this is a very conservative estimate a better accuracy might be 
needed for future experiments; higher luminosity $B$-factories or even 
the International Linear Collider (ILC). Two loop corrections, at least 
in the leading log approximation, will reduce this 
uncertainty to at least 1 to 2 per mil.

\section{Interplay between $e^+e^-$ and tau data}

\begin{table}
\caption{Contribution of the pion form factor to $a_\mu^{\rm{had, LO}}$ 
(in units of $10^{-11}$) from the different energy regions. 
Numbers from Ref.~\cite{eidelman}($^*$ my own estimate).}
\begin{center}
\begin{tabular}{lc} \hline\hline
390-520 MeV & \\
SND &   $478.0 \pm 17.3 \pm 6.9 (18.6)$ \\
CMD-2 & $461.7 \pm  9.8 \pm 3.2 (10.3)$ \\ \hline
520-600 MeV & \\
SND & $425.0 (27)^*$ \\ \hline
600-960 MeV & \\
SND &   $3768 \pm 13 \pm 47 (48)$ \\
CMD-2 & $3771 \pm 19 \pm 27 (33)$ \\
KLOE &  $3756 \pm  8 \pm 49 (50)$ \\ \hline\hline
\end{tabular}
\end{center}
\end{table}

The hadronic vacuum polarization contribution to the SM prediction of the
anomalous magnetic moment of the muon is obtained though a dispersion 
integral over the $e^+e^-$ hadronic cross section (once ISR and vacuum 
polarization corrections are subtracted) 
normalized to the point-like muonic cross section $\sigma=4\pi\alpha^2/3s$,
where the low energy data has the largest weight.
This requires a complicated unfolding of radiative corrections. 
Alternatively, conservation of vector currents (CVC) and isospin symmetry 
allows to use the hadronic vector spectral function of tau lepton decays 
as input to this dispersion integral.    
At present, the SM prediction obtained from tau and $e^+e^-$ data do not 
agree to each other. This might be an indication of unaccounted isospin 
breaking contributions~\cite{lopezcastro} but it is at the same time 
rather puzzling because precisely the tau based analysis 
agrees with the measurement of the anomalous magnetic moment from the 
BNL experiment~\cite{hertzog}, while the $e^+e^-$ based analysis doesn't. 
This discrepancy is of the order of three standard deviations~\cite{davier}.
Recent preliminary BELLE~\cite{fujikawa} tau data seem to point 
however towards a better agreement between $e^+e^-$ and tau data. 
The branching ratio of $\tau \to \nu_\tau \pi^+ \pi^0$ calculated from 
$e^+e^-$ data neither agrees~\cite{davier}. 

The situation among $e^+e^-$ data is also puzzling. The integral over 
the energy scan (SND~\cite{SND} and CMD-2~\cite{CMD2}) and 
the radiative return (KLOE~\cite{KLOElarge}) data, and 
therefore the prediction for $a_\mu$, agrees within $0.5 \sigma$.
This agreement is due to a tricky compensation of discrepancies 
in the shape of the energy distribution above and below the 
$\rho$-peak. 

A careful regard to these data tell us that the three experiments
do overlap indeed only in the energy region above $600$~MeV,
while in the threshold region only data from the Novosibirsk 
experiments are available. 
As expected the statistical error of radiative return data is much 
better than of energy scan. CMD-2 has the better systematic error 
($0.6\%$), while the total systematic error of SND and KLOE are 
comparable ($1.3\%$). The new KLOE analysis~\cite{venanzoni} 
will however reduce it to less than $1\%$. 
While SND and CMD-2 experiments agree to 
each other in the full energy range, they can not be considered completely 
independent as both analysis use the same radiative correction package. 
Furthermore, the new CMD-2 data do not cover the region between 
$520$ and $600$~MeV. This makes the ongoing KLOE large photon polar 
angle analysis particularly interesting because the agreement 
of the integral is very unlikely to happen in the threshold region.
Excluding a set of data from the prediction of $a_\mu$, until these 
discrepancies are solved, might lead to a biased result.

\section{Summary}

The radiative return has proven to be a competitive method for 
the precise measurement of the hadronic cross section, detailed 
studies of hadronic interactions, and even discoveries of new 
resonances. Many interesting problems, for example
a proper modeling of the hadronic current of multi-meson 
final states, FSR simulation for more than two-pions,
modeling of narrow resonances and many others not mentioned
await still for detailed theoretical investigations.

New data from the KLOE experiment at small polar photon angles 
with a total systematic uncertainty below $1\%$ and in particular
in the threshold region from the large photon polar angle 
analysis~\cite{venanzoni}, the long awaited pion form factor 
measurement at $B$-factories, but also future new data from 
the energy scan~\cite{eidelman} should help to clarify 
the discrepancies between different $e^+e^-$ data sets.
To a great extent independent test of radiative corrections 
are also needed. One should remember that even the direct 
measurement of the ratio $\sigma_{\pi\pi}/\sigma_{\mu\mu}$,
where most systematics are expected to cancel, requires a
careful treatment of radiative corrections and unfolding, 
due to the fact that radiative corrections affect differently 
the angular distributions of pions and muons~\cite{Rodrigo:2001kf}, 
and because the $R$-ratio entering the dispersion integral 
for $a_\mu$ is normalized to the point-like cross section 
of the muon but not to the physical cross section. 
All together is necessary to confirm 
the discrepancy of the $(g-2)_\mu$ measurement with respect
to the Standard Model prediction. Understanding tau data,
whether the agreement with $(g-2)_\mu$ is by accident or not,  
is also crucial for this purpose.

\section*{Acknowledgments}

It is a pleasure to thank H.~Czy\.z, A.~Grzeli\'nska, J.~H.~K\"uhn and 
E.~Nowak for a very fruifull collaboration.


\begin{thebibliography}{9}

\bibitem{KLOElarge}
  A.~Aloisio {\it et al.}  [KLOE Collaboration],
  Phys.\ Lett.\ B {\bf 606} (2005) 12.

\bibitem{Aubert:2004kj}
  B.~Aubert {\it et al.}  [BABAR Collaboration],
  Phys.\ Rev.\ D {\bf 70} (2004) 072004.

\bibitem{wang}
W.F.~Wang, these proceedings.



\bibitem{Binner:1999bt}
  S.~Binner, J.~H.~K\"uhn and K.~Melnikov,
  Phys.\ Lett.\ B {\bf 459} (1999) 279.
  H.~Czy\.z and J.~H.~K\"uhn,
  Eur.\ Phys.\ J.\ C {\bf 18} (2001) 497.


\bibitem{Rodrigo:2001kf}
  G.~Rodrigo, H.~Czy\.z, J.~H.~K\"uhn and M.~Szopa,
  Eur.\ Phys.\ J.\ C {\bf 24} (2002) 71.
  G.~Rodrigo, A.~Gehrmann-De Ridder, M.~Guilleaume and J.~H.~K\"uhn,
  Eur.\ Phys.\ J.\ C {\bf 22} (2001) 81.

\bibitem{Kuhn:2002xg}
  J.~H.~K\"uhn and G.~Rodrigo,
  Eur.\ Phys.\ J.\ C {\bf 25} (2002) 215.
  G.~Rodrigo,
  Acta Phys.\ Polon.\ B {\bf 32} (2001) 3833.

\bibitem{Czyz:2002np}
 H.~Czy\.z, A.~Grzeli\'nska, J.~H.~K\"uhn and G.~Rodrigo,
  Eur.\ Phys.\ J.\ C {\bf 27} (2003) 563.

\bibitem{Czyz:2003ue}
 H.~Czy\.z, A.~Grzeli\'nska, J.~H.~K\"uhn and G.~Rodrigo,
  Eur.\ Phys.\ J.\ C {\bf 33} (2004) 333;
  Eur.\ Phys.\ J.\ C {\bf 39} (2005) 411.

\bibitem{Czyz:2005as}
  H.~Czy\.z, A.~Grzeli\'nska, J.~H.~K\"uhn and G.~Rodrigo,
  Eur.\ Phys.\ J.\ C {\bf 47} (2006) 617.

\bibitem{Czyz:2004ua}
  H.~Czy\.z, J.~H.~K\"uhn, E.~Nowak and G.~Rodrigo,
  Eur.\ Phys.\ J.\ C {\bf 35} (2004) 527.


\bibitem{Jadach:1999vf}
  S.~Jadach, B.~F.~L.~Ward and Z.~Was,
  Comput.\ Phys.\ Commun.\  {\bf 130} (2000) 260.

\bibitem{Jadach:KKMC}
  S.~Jadach, B.~F.~L.~Ward and S.~A.~Yost,
  Phys.\ Rev.\ D {\bf 73} (2006) 073001.
  C.~Glosser, S.~Jadach, B.~F.~L.~Ward and S.~A.~Yost,
  Phys.\ Lett.\ B {\bf 605} (2005) 123.

\bibitem{venanzoni}
G.~Venanzoni, these proceedings.


\bibitem{Hoefer:2001mx}
  A.~H\"ofer, J.~Gluza and F.~Jegerlehner,
  Eur.\ Phys.\ J.\ C {\bf 24} (2002) 51.

\bibitem{Czyz:2006dm}
  H.~Czy\.z and E.~Nowak-Kubat,
  Phys.\ Lett.\ B {\bf 634}, 493 (2006).

\bibitem{Melnikov:2000gs}
  K.~Melnikov, F.~Nguyen, B.~Valeriani and G.~Venanzoni,
  Phys.\ Lett.\ B {\bf 477} (2000) 114.


\bibitem{Czyz:2004nq}
  H.~Czy\.z, A.~Grzeli\'nska and J.~H.~K\"uhn,
  Phys.\ Lett.\ B {\bf 611} (2005) 116.

\bibitem{KLOEphi}
  F.~Ambrosino {\it et al.}  [KLOE Collaboration],
  Phys.\ Lett.\ B {\bf 634} (2006) 148.

\bibitem{Bramon:1992ew}
  A.~Bramon, G.~Colangelo, P.~J.~Franzini and M.~Greco,
  Phys.\ Lett.\ B {\bf 287}, 263 (1992).

\bibitem{LucioMartinez:1994yu}
  J.~L.~Lucio Martinez and M.~Napsuciale,
  Phys.\ Lett.\ B {\bf 331}, 418 (1994).

\bibitem{Pancheri:2006cp}
  G.~Pancheri, O.~Shekhovtsova and G.~Venanzoni,
  Phys.\ Lett.\ B {\bf 642} (2006) 342.

\bibitem{Aubert:2005cb}
  B.~Aubert {\it et al.}  [BABAR Collaboration],
  Phys.\ Rev.\ D {\bf 73} (2006) 012005.


\bibitem{Aubert:2005rm}
  B.~Aubert {\it et al.}  [BABAR Collaboration],
  Phys.\ Rev.\ Lett.\  {\bf 95} (2005) 142001.

\bibitem{Coan:2006rv}
  T.~E.~Coan {\it et al.}  [CLEO Collaboration],
  Phys.\ Rev.\ Lett.\  {\bf 96} (2006) 162003.

\bibitem{He:2006kg}
  Q.~He  [CLEO Collaboration],
  Phys.\ Rev.\ D {\bf 74} (2006) 091104.

\bibitem{Abe:2006hf}
  K.~Abe  [Belle Collaboration],
  hep-ex/0612006.

\bibitem{lopezcastro}
G.~L\'opez Castro, these proceedings;
  A.~Flores-Tlalpa {\it et al.}, 
  hep-ph/0611226.

\bibitem{hertzog}
D.W.~Hertzog, these proceedings.

\bibitem{davier}
M.~Davier, these proceedings.

\bibitem{fujikawa}
M.~Fujikawa, these proceedings.



\bibitem{SND}
  M.~N.~Achasov {\it et al.} [SND Collaboration],
  J.\ Exp.\ Theor.\ Phys.\  {\bf 103} (2006) 380.

\bibitem{CMD2}
  R.~R.~Akhmetshin {\it et al.} [CMD-2 Collaboration],
  JETP Lett.\  {\bf 84} (2006) 413.
  Phys.\ Lett.\ B {\bf 578} (2004) 285.


\bibitem{eidelman}
S.I.~Eidelman, these proceedings.









\end{thebibliography}
\end{document}